\documentclass[11pt]{article}
\usepackage{graphicx}
\usepackage{amsmath}
\usepackage{amssymb}
\usepackage{amsxtra}

\title{The renormalization group invariants and exact results for various supersymmetric theories}
\author{K.V. Stepanyantz\\Moscow State University, 119991, Moscow, Russia,\\ stepan@m9com.ru}

\begin{document}
\maketitle

\begin{abstract}
Some recent all-loop results on the renormalization of supersymmetric theories are summarized and reviewed. In particular, we discuss how it is possible to construct expressions which do not receive quantum corrections in all orders for certain ${\cal N}=1$ supersymmetric theories. For instance, in ${\cal N}=1$ SQED+SQCD there is a renormalization group invariant combination of two gauge couplings. For the Minimal Supersymmetric Standard Model there are two such independent combinations of the gauge and Yukawa couplings. We investigate the scheme-dependence of these results and verify them by explicit three-loop calculations. We also argue that the all-loop exact $\beta$-function and the corresponding renormalization group invariant can exist in the $6D$, ${\cal N}=(1,0)$ supersymmetric higher derivative gauge theory interacting with a hypermultiplet in the adjoint representation.
\end{abstract}

\section{Introduction}
\hspace*{\parindent}

Quantum corrections can tell a lot about the structure of the surrounding world. For instance, comparing the theoretical predictions for the electron anomalous magnetic moment with the experimental data we conclude that it is necessary to describe nature by quantum field theory \cite{Peskin:1995ev}. Analyzing the unification of the running gauge couplings in the Standard model and its supersymmetric extensions \cite{Ellis:1990wk,Amaldi:1991cn,Langacker:1991an} and comparing the results with the predictions of the Grand Unified theories we obtain a strong indirect indication for the existence of supersymmetry \cite{Mohapatra:1986uf,Raby:2017ucc}. One more evidence in favour of supersymmetry in the high energy physics is provided by the absence of quadratically divergent quantum corrections to the Higgs boson mass in this case. Moreover, the detailed analysis of quantum corrections to the (lightest) Higgs boson mass, to the anomalous magnetic moment of muon, etc. can provide some important information about new physics beyond the Standard model, see the reviews in \cite{ParticleDataGroup:2024cfk} and references therein. Some important information can also be obtained by investigating the renormalization group invariants (RGIs), which, by definition, are the scale independent values. Some of them can be approximate (see, e.g., \cite{Raby:2017ucc,Jack:1996qq}), but sometimes it is even possible to construct the expressions that are scale independent in all orders (see, e.g. \cite{Novikov:1983uc,Hisano:1997ua}).

In this paper we describe how one can construct all-order exact RGIs in certain supersymmetric theories. Namely, we consider ${\cal N}=1$ supersymmetric chromodynamics interacting with supersymmetric electrodynamics (SQCD+SQED), the Mininmal Supersymmetric Standard Model (MSSM), and the higher derivative $6D$, ${\cal N}=(1,0)$ supersymmetric Yang--Mills theory theory interacting with the hypermultiplet in the adjoint representation. For all these theories we will construct such combinations of couplings that do not depend on scale in all orders of the perturbation theory.

The paper is organized as follows. In Sect. \ref{Section_Superspace} some basic information about the superfield formulation of the supersymetric theories is recalled. Some features of quantum corrections in supersymmetric theories that are revealed by the higher covariant regularization are discussed in Sect. \ref{Section_Higher_Derivatives}. After that, in Sect. \ref{Section_SQCD+SQED} the RGI composed from two gauge couplings is constructed for ${\cal N}=1$ SQCD+SQED. Analogous RGIs for MSSM are presented in Sect. \ref{Section_MSSM}. A possibility of constructing an RGI for a certain $6D$, ${\cal N}=(1,0)$ higher derivative supersymmetric theory is analyzed in Sect. \ref{Section_Harmonic_Superspace}. A brief summary of the results is given in Conclusion.

\section{Superfield formultion of supersymmetric theories and some aspects of their renormalization}
\hspace*{\parindent}\label{Section_Superspace}

In ${\cal N}=1$ superspace renormalizable supersymmetric gauge theories (with a single gauge coupling) are described by the manifestly supersymmetric action
\begin{eqnarray}\label{Action_General}
&& S = \frac{1}{2 e_0^2}\,\mbox{Re}\,\mbox{tr}\int d^4x\, d^2\theta\,W^a W_a + \frac{1}{4} \int d^4x\, d^4\theta\,\phi^{*i}
(e^{2V})_i{}^j \phi_j\nonumber\\
&& + \Big\{ \int d^4x\,d^2\theta\, \Big(\frac{1}{4} m_0^{ij}\phi_i\phi_j + \frac{1}{6}\lambda_0^{ijk} \phi_i \phi_j \phi_k \Big) + \mbox{c.c.}\Big\}.\qquad
\end{eqnarray}

\noindent
Here $\phi_i$ are the chiral matter superfields in the representation $R$ of the gauge group $G$, which, by definition, satisfy the constraint $\bar D_{\dot a}\phi_i=0$. (In our notation $\bar D_{\dot a}$ and $D_a$ denote the left and right components of the supersymmetric covariant derivative, respectively.) The gauge superfield $V$ is Hermitian, $V^+ = V$. Its gauge field strength is the chiral superfield
\begin{equation}
W_a = \frac{1}{8} \bar D^2 \Big(e^{-2V} D_a e^{2V}\Big).
\end{equation}

The renormalizability requires that the superpotential should be no more than cubic in chiral superfields. According to the well-known theorem \cite{Grisaru:1979wc}, in this case it does not receive divergent quantum corrections. However, nonrenormalization of the superpotential does not imply that masses and Yukawa couplings in the supersymmetric case are not renormalized. In fact, their renormalization appears to be related to the renormalization of the chiral matter superfields. Namely, if $\phi_i = (\sqrt{Z})_i{}^j \phi_{R,j}$, then
\begin{equation}
m^{ij} = m_0^{kl} (\sqrt{Z})_k{}^i (\sqrt{Z})_l{}^j; \qquad \lambda^{ijk} = \lambda_0^{mnp} (\sqrt{Z})_m{}^i (\sqrt{Z})_n{}^j (\sqrt{Z})_p{}^k.
\end{equation}

Unexpectedly, the renormalization of the gauge couplings in the supersymmetric case is also related to the renormalization of chiral matter superfields by the exact Novikov, Shifman, Vainshtein, and Zakharov (NSVZ) $\beta$-function \cite{Novikov:1983uc,Jones:1983ip,Novikov:1985rd,Shifman:1986zi}. For theories with a single gauge coupling it can be written in the form
\begin{equation}\label{Beta_NSVZ}
\beta(\alpha,\lambda) = - \frac{\alpha^2\Big(3 C_2 - T(R) + C(R)_i{}^j (\gamma_\phi)_j{}^i(\alpha,\lambda)/r\Big)}{2\pi(1-C_2\alpha/2\pi)},
\end{equation}

\noindent
where $\alpha = e^2/4\pi$ and $\lambda$ are the gauge and Yukawa coupling constants, respectively, and $r\equiv \mbox{dim}\,G$. For the pure ${\cal N}=1$ SYM theory (which does not contain chiral matter superfields) the NSVZ equation produces the all-order exact formula for the $\beta$-function, which can equivalently be reformulated as the statement that in all orders of the perturbation theory the expression
\begin{equation}\label{N=1_SYM_RGI}
\Big(\frac{\mu^3}{\alpha}\Big)^{C_2} \exp\Big(-\frac{2\pi}{\alpha}\Big) = \mbox{RGI}
\end{equation}

\noindent
does not receive quantum corrections \cite{Novikov:1985rd}.

Nevertheless, it is necessary to remember that the all-loop equations describing the renormalization of supersymmetric theories are valid only for certain renormalization prescriptions. For instance, in the $\overline{\mbox{DR}}$-scheme the NSVZ equation does not hold starting from the order $O(\alpha^4)$ (the three-loop approximation for the $\beta$-function and the two-loop approximation for the anomalous dimension) \cite{Jack:1996vg,Jack:1996cn,Jack:1998uj,Harlander:2006xq}. However, the all-loop NSVZ renormalization prescription can be constructed with the help of the higher covariant derivative regularization \cite{Slavnov:1971aw,Slavnov:1972sq,Slavnov:1977zf} in the superfield formulation \cite{Krivoshchekov:1978xg,West:1985jx} considered in the next section.

\section{Quantum properties of supersymmetric theories regularized by higher covariant derivatives}
\hspace*{\parindent}\label{Section_Higher_Derivatives}

For supersymmetric theories the higher covaraint derivative regularization allows revealing some nontrivial feature of quantum corrections which are not seen in the case of using the dimensional reduction, see, e.g., \cite{Stepanyantz:2017sqg,Stepanyantz:2019lyo}. It is introduced by adding to the action (\ref{Action_General}) terms with higher gauge and supersymmetric covariant derivatives (denoted by $\bar\nabla_{\dot a}$ and $\nabla_a$). After that, the regularized action takes the form
\begin{eqnarray}\label{Action_Regularized}
&&\hspace*{-5mm} S_{\mbox{\tiny reg}} = \frac{1}{2 e_0^2}\,\mbox{Re}\, \mbox{tr} \int d^4x\, d^2\theta\, W^a \Big[e^{-2V} R\Big(-\frac{\bar\nabla^2 \nabla^2}{16\Lambda^2}\Big) e^{2V}\Big]_{Adj} W_a \nonumber\\
&&\hspace*{-5mm}  + \frac{1}{4} \int d^4x\,d^4\theta\, \phi^{*i} \Big[F\Big(-\frac{\bar\nabla^2 \nabla^2}{16\Lambda^2}\Big) e^{2V}\Big]_i{}^j \phi_j
\nonumber\\
&&\hspace*{-5mm}
+ \Big[\, \int d^4x\,d^2\theta\, \Big(\frac{1}{4} m_0^{ij} \phi_i \phi_j + \frac{1}{6} \lambda_0^{ijk}\phi_i \phi_j \phi_k \Big) + \mbox{c.c.} \Big],
\end{eqnarray}

\noindent
where the functions $R(x)$ and $F(x)$ should rapidly increase at infinity and satisfy the condition $R(0)=F(0)=1$. Due to the former property of these functions, divergences remain only in the one-loop approximation. For regularizing these residual divergences, one can insert into the generating functional some special Pauli--Villars determinants \cite{Slavnov:1977zf}. The details of the corresponding construction can be found in \cite{Aleshin:2016yvj,Kazantsev:2017fdc}.

As was noted in numerous calculations made in supersymmetric theories in the lowest (up to four loops) orders of the perturbation theory with the higher derivative regularization \cite{Smilga:2004zr,Pimenov:2009hv,Buchbinder:2014wra,Buchbinder:2015eva,Shakhmanov:2017wji,Kazantsev:2018nbl,Kuzmichev:2019ywn,Shirokov:2023jya}, the NSVZ equation appears in this case because the integrals giving the $\beta$-function defined in terms of the bare couplings are integrals of double total derivatives with respect to the loop momenta. For instance, in the one-loop approximation the $\beta$-function for the theory (\ref{Action_General}) is written in the form \cite{Aleshin:2016yvj}
\begin{eqnarray}
&&\hspace*{-7mm} \frac{\beta(\alpha_0,\lambda_0)}{\alpha_0^2}
=  \int \frac{d^4q}{(2\pi)^4} \frac{d}{d\ln\Lambda}
\frac{\partial}{\partial q^\mu} \frac{\partial}{\partial q_\mu}
\bigg\{-\frac{\pi C_2}{q^2}\bigg[ \ln\Big(1 + \frac{M_\varphi^2}{q^2 R^2(q^2/\Lambda^2)}\Big)
\nonumber\\
&&\hspace*{-7mm} + 2 \ln\Big(1+ \frac{M_\varphi^2}{q^2}\Big)\bigg] + \frac{\pi T(R)}{q^2} \ln\Big(1 + \frac{M^2}{q^2
F^2(q^2/\Lambda^2)}\Big)\bigg\}
+ O(\alpha_0,\lambda_0^2),
\end{eqnarray}

\noindent
where $M$ and $M_\varphi$ are masses of the Pauli--Villars superfields, and a small vicinity of the singular point $q^\mu=0$ should be excluded from the integration region. In all orders the factorization of integrals giving the $\beta$-function into integrals of double total derivatives has been proved in \cite{Stepanyantz:2011jy} in the Abelian case and in \cite{Stepanyantz:2019ihw} for general non-Abelian gauge theories.

The double total derivatives effectively cut internal lines in (vacuum) supergraphs thus reducing a number of loop integrations by 1 and giving superdiagrams contributing to the anomalous dimensions of various quantum superfields. After that, (for ${\cal N}=1$ supersymmetric theories regularized by higher covarant derivatives) the NSVZ $\beta$-function is obtained in all orders for the renormalization group functions (RGFs) defined in terms of the bare couplings by summing singular contributions \cite{Stepanyantz:2020uke} and taking into account the nonrenormalization of the triple gauge-ghost vertices \cite{Stepanyantz:2016gtk}.

For the standard RGFs defined in terms of the renormalized couplings the NSVZ equation holds in all orders in the HD+MSL scheme \cite{Kataev:2013eta}, when a theory is regularized by Higher Derivatives, and divergences are removed by Minimal Subtractions of Logarithms \cite{Stepanyantz:2017sqg,Shakhmanov:2017wji}. This in particular implies that for the pure ${\cal N}=1$ SYM theory the RGI (\ref{N=1_SYM_RGI}) is valid in the HD+MSL scheme and is not valid in the $\overline{\mbox{DR}}$ scheme.

Involving the statement that the HD+MSL is an all-loop NSVZ scheme, it is possible to use the NSVZ equation for obtaining the $\beta$-function(s) in higher orders on the base of the anomalous dimension(s) in the previous loops, see, e.g., \cite{Kazantsev:2020kfl,Haneychuk:2022qvu,Shirokov:2022jyd,Haneychuk:2025ehb}.

\section{Renormalization of ${\cal N}=1$ SQCD+SQED }
\hspace*{\parindent}\label{Section_SQCD+SQED}

Following \cite{Kataev:2024amm}, we argue that in ${\cal N}=1$ SQCD+SQED one can construct an all-loop RGI from two gauge couplings $\alpha_s \equiv g^2/4\pi$ and $\alpha=e^2/4\pi$. In the massless limit this theory is described by the superfield action
\begin{eqnarray}\label{SQCD_Action}
&& S = \frac{1}{2g^2}\,\mbox{Re}\,\mbox{tr}\int d^4x\,d^2\theta\, W^a W_a + \frac{1}{4e^2}\,\mbox{Re}\int d^4x\, d^2\theta\, \mbox{\boldmath$W$}^a \mbox{\boldmath$W$}_a\qquad\nonumber\\
&& + \sum\limits_{\mbox{\scriptsize a}=1}^{N_f}\, \frac{1}{4}\int d^4x\, d^4\theta\, \Big(\phi_{\mbox{\scriptsize a}}^+ e^{2V + 2q_{\mbox{\scriptsize a}}\mbox{\scriptsize \boldmath$V$}}\phi_{\mbox{\scriptsize a}}
+ \widetilde\phi_{\mbox{\scriptsize a}}^+ e^{-2V^T - 2q_{\mbox{\scriptsize a}}\mbox{\scriptsize \boldmath$V$}} \widetilde\phi_{\mbox{\scriptsize a}}\Big),\qquad
\end{eqnarray}

\noindent
where the subscript $\mbox{a}$ numerates flavors. It is invariant under the transformations of the group $G\times U(1)$, $V$ and $\mbox{\boldmath$V$}$ being the gauge superfields corresponding to the subgroups $G$ and $U(1)$, respectively.
The chiral matter superfields $\phi_{\mbox{\scriptsize a}}$ and $\widetilde\phi_{\mbox{\scriptsize a}}$ belong to the (conjugated) representations $R$ and $\bar R$, respectively, and have opposite $U(1)$ charges $\pm q_{\mbox{\scriptsize a}} e$.

For the theory (\ref{SQCD_Action}) the renormalization of the gauge couplings is described by the NSVZ $\beta$-functions, which are also valid for theories with multiple gauge couplings, see \cite{Shifman:1996iy,Korneev:2021zdz}. For an irreducible representation $R$ and $q_{\mbox{\scriptsize a}} = 1$ they are written as
\begin{eqnarray}\label{SQCD_Betas}
&& \frac{\beta_s(\alpha_s,\alpha)}{\alpha_s^2} = - \frac{1}{2\pi(1-C_{2} \alpha_s/2\pi)} \bigg[\, 3 C_2 - 2 T(R) N_f \Big(1-\gamma(\alpha_s,\alpha)\Big) \bigg];\qquad\nonumber\\
&& \frac{\beta(\alpha,\alpha_s)}{\alpha^2} = \frac{1}{\pi}\, \mbox{dim}\,R\, N_f \Big(1-\gamma(\alpha_s,\alpha)\Big),
\end{eqnarray}

\noindent
where we took into account that all anomalous dimensions coincide in the particular case under consideration.  Comparing two expressions in Eq. (\ref{SQCD_Betas}), we see that the anomalous dimension of the matter superfeilds can be eliminated, and the gauge $\beta$-functions satisfy the all-order exact equation
\begin{equation}\label{SQCD_Beta_Relation}
\Big(1-\frac{C_{2}\alpha_s}{2\pi}\Big) \frac{\beta_s(\alpha_s,\alpha)}{\alpha_s^2} = - \frac{3 C_2}{2\pi} + \frac{T(R)}{\mbox{dim}\,R}\cdot \frac{\beta(\alpha,\alpha_s)}{\alpha^2},
\end{equation}

\noindent
which relates running of the strong and electromagnetic couplings in the theory under consideration. Evidently, Eq. (\ref{SQCD_Beta_Relation}) should be valid in the HD+MSL scheme, because the original NSVZ equations are also satisfied for this renormalization prescription. Integrating Eq. (\ref{SQCD_Beta_Relation}) we easily obtain that the expression
\begin{equation}\label{SQCD+SQED_RGI}
\Big(\frac{\alpha_s}{\mu^3}\Big)^{C_2} \exp\Big(\frac{2\pi}{\alpha_s} - \frac{T(R)}{\mbox{dim}\,R}\cdot \frac{2\pi}{\alpha}\Big)= \mbox{RGI}
\end{equation}

\noindent
vanishes after differentiating with respect to $\ln\mu$. This implies that it does not depend on scale and, therefore, is an RGI.

For the theory (\ref{SQCD_Action}) with different $U(1)$ charges $q_{\mbox{\scriptsize a}}$ Eq. (\ref{SQCD_Beta_Relation}) does not hold. However, in this case it is possible to relate the ${\cal N}=1$ SQCD $\beta$-function and the Adler $D$-function \cite{Adler:1974gd} in all orders by the equation \cite{Kataev:2024amm},
\begin{equation}\label{Beta_Ader_Equation}
\beta_s(\alpha_s) = - \frac{\alpha_s^2}{2\pi(1-C_{2} \alpha_s/2\pi)} \bigg[\, 3 C_2 - \frac{4\, T(R) N_f D(\alpha_s)}{3\,\mbox{\boldmath$q^2$}\, \mbox{dim}\,R }  \bigg],
\end{equation}

\noindent
where ${\displaystyle \mbox{\boldmath$q^2$}} \equiv \sum\limits_{\mbox{\scriptsize a}=1}^{N_f} (q_{\mbox{\scriptsize a}})^2$.

In the the three-loop approximation (where the scheme dependence becomes essential) Eqs. (\ref{SQCD_Beta_Relation}) and (\ref{Beta_Ader_Equation}) have been verified by explicit calculations in \cite{Haneychuk:2025ehb}, where the three-loop $\beta$-functions for the ${\cal N}=1$ SQCD+SQED have been calculated for a general renormalization prescription supplementing the higher covariant derivative regularization.  In particular, it has been demonstrated that in the HD+MSL scheme these equations are satisfied this approximation and are not valid  in the $\overline{\mbox{DR}}$ scheme. This in particular implies that the expression (\ref{SQCD+SQED_RGI}) is not an RGI in the $\overline{\mbox{DR}}$ scheme starting from three loops, where the scheme dependence manifests itself.

\section{The Minimal Supersymmetric Standard Model}
\hspace*{\parindent}\label{Section_MSSM}

The MSSM is the simplest supersymmetric extension of the Standard Model. It is a gauge theory with the group $SU(3)\times SU(2)\times U(1)$ and softly broken supersymmetry, where quarks, leptons, and Higgs fields are components of the chiral matter superfields. Evidently, in the MSSM there are 3 gauge couplings
\begin{equation}
\alpha_3 = \frac{e_3^2}{4\pi};\qquad \alpha_2 = \frac{e_2^2}{4\pi};\qquad  \alpha_1 = \frac{5}{3}\cdot \frac{e_1^2}{4\pi}
\end{equation}

\noindent
corresponding to the subgroups $SU(3)$, $SU(2)$, and $U(1)$, respectively. The MSSM action also contains dimensionless Yukawa couplings $(Y_U)_{IJ}$, $(Y_D)_{IJ}$, and $(Y_E)_{IJ}$ (which are $3\times 3$ matrices) and the parameter $\mbox{\boldmath$\mu$}$ with the dimension of mass inside the superpotential
\begin{eqnarray}
&&\hspace*{-3mm} W = \left(Y_U\right)_{IJ}
\left(\widetilde U\ \widetilde D \right)^{a}_I
\left(
\begin{array}{cc}
0 & 1\\
-1 & 0
\end{array}
\right)
\left(
\begin{array}{c}
H_{u1}\\ H_{u2}
\end{array}
\right) U_{aJ} + \left(Y_D\right)_{IJ}
\left(\widetilde U\ \widetilde D \right)^{a}_I
\nonumber\\
&&\hspace*{-3mm} \times \left(
\begin{array}{cc}
0 & 1\\
-1 & 0
\end{array}
\right)
\left(
\begin{array}{c}
H_{d1}\\ H_{d2}
\end{array}
\right) D_{aJ}
+ \left(Y_E\right)_{IJ} \left(\widetilde N\ \widetilde E \right)_{I}
\left(
\begin{array}{cc}
0 & 1\\
-1 & 0
\end{array}
\right)
\left(
\begin{array}{c}
H_{d1}\\ H_{d2}
\end{array}
\right)\nonumber\\
&&\hspace*{-3mm} \times E_J
+ \mbox{\boldmath$\mu$} \left(H_{u1}\ H_{u2} \right)
\left(
\begin{array}{cc}
0 & 1\\
-1 & 0
\end{array}
\right)
\left(
\begin{array}{c}
H_{d1}\\ H_{d2}
\end{array}
\right).
\end{eqnarray}

The renormalization group running of the gauge couplings in the MSSM ia described exactly in all loops by three NSVZ equations \cite{Shifman:1996iy}, which relate three gauge $\beta$-functions of the theory to the anomalous dimensions of the chiral matter superfields. Similarly, RGFs describing the renormalization of the Yukawa couplings and of the parameter $\mbox{\boldmath$\mu$}$ are also related to the anomalous dimensions of the matter superfields due to the nonrenormalization of the superpotential \cite{Grisaru:1979wc}.

According to \cite{Rystsov:2024soq}, after eliminating the anomalous dimensions of the chiral matter superfields and $\mbox{\boldmath$\mu$}$ from the resulting system of (all-order exact in the HD+MSL scheme) equations we obtain
\begin{eqnarray}\label{Invariance_Equations}
&&\hspace*{-4mm} 0 = \Big(\frac{1}{\alpha_2} - \frac{\pi}{\alpha_2^2}\Big) \beta_2 - \frac{5\pi}{3\alpha_1^2} \beta_1 + 6 + 3\gamma_{\mbox{\scriptsize \boldmath$\mu$}} - \gamma_{\mbox{\tiny det}\,Y_E} - \frac{4}{3}\gamma_{\mbox{\tiny det}\,Y_U} - \frac{1}{3} \gamma_{\mbox{\tiny det}\,Y_D};\qquad\nonumber\\
&&\hspace*{-4mm} 0 = \Big(\frac{3}{\alpha_3} - \frac{2\pi}{\alpha_3^2}\Big) \beta_3 -3 + 3\gamma_{\mbox{\scriptsize \boldmath$\mu$}} - \gamma_{\mbox{\tiny det}\,Y_U} - \gamma_{\mbox{\tiny det}\,Y_D}.
\end{eqnarray}

\noindent
Integrating these equations we obtain that the expressions
\begin{eqnarray}\label{RGIs}
&&\hspace*{-7mm} \mbox{RGI}_1\equiv \frac{\mbox{\boldmath$\mu$}^{3}\,\mu^6\,\alpha_2}{\big(\mbox{det}\, Y_E\big)\,\big(\mbox{det}\, Y_U\big)^{4/3}\, \big(\mbox{det}\, Y_D\big)^{1/3}}\, \exp\Big(\frac{\pi}{\alpha_2} +\frac{5\pi}{3\alpha_1} \Big);\nonumber\\
&&\hspace*{-7mm} \mbox{RGI}_2\equiv \frac{\mbox{\boldmath$\mu$}^{3}\,(\alpha_3)^3}{\mu^3\,\mbox{det}\, Y_U\, \mbox{det}\, Y_D}\, \exp\Big(\frac{2\pi}{\alpha_3} \Big)
\end{eqnarray}

\noindent
do not depend on the renormalization scale {\it in all orders}. However, this renormalization group invariance is valid only for some special renormalization prescriptions. In fact, the scheme dependence of the equations becomes essential starting from the order $O(\alpha^2,\alpha Y^2,Y^4)$ corresponding to the three-loop approximation for the $\beta$-functions and to the two-loop approximation for the anomalous dimensions. In the HD+MSL scheme these RGFs have been calculated in \cite{Haneychuk:2022qvu}. After substituting them into Eq. (\ref{Invariance_Equations}) it was demonstrated that
in the HD+MSL scheme these equation are really satisfied independently of the values of regularization parameters. However, in the $\overline{\mbox{DR}}$ scheme the expressions in the right hand sides of Eq. (\ref{Invariance_Equations}) do not vanish in that orders where the scheme dependence becomes essential. (The corresponding RGFs needed for making this verification were taken from \cite{Jack:2004ch}.) Therefore, in this scheme the expressions (\ref{RGIs}) are scale independent only in the two first orders of the perturbation theory.

\section{$6D$, ${\cal N}=(1,0)$ higher derivative theory in the harmonic superspace}
\hspace*{\parindent}\label{Section_Harmonic_Superspace}

It would be interesting to reveal if quantum corrections in supersymmetric theories may remain their attractive features in higher dimensions. Usual supersymmetric theories in higher dimensions are not renormalizable, because the degree of divergence increases with the number of loops. However, in this case it is possible to consider theories with higher derivatives. It is convenient to describe them using $6D$, ${\cal N}=(1,0)$ harmonic superspace \cite{Howe:1983fr,Howe:1985ar,Zupnik:1986da,Ivanov:2005kz,Buchbinder:2014sna} analogous to the usual $4D$, ${\cal N}=2$ harmonic superspace \cite{Galperin:1984av,Galperin:2001uw}, because in this case ${\cal N}=(1,0)$ supersymmetry is manifest.

Following Ref. \cite{Buchbinder:2025aac}, we consider the $6D$, ${\cal N}=(1,0)$ supersymmetric theory similar to the one presented in \cite{Ivanov:2005qf}, which in the harmonic superspace is described by the action
\begin{equation}\label{6D_Action}
S = \pm \frac{1}{2e_0^2} \mbox{tr} \int d\zeta^{(-4)} (F^{++})^2 - \frac{2}{e_0^2}\,\mbox{tr} \int d\zeta^{(-4)} \widetilde{q^+} \nabla^{++} q^+.
\end{equation}

\noindent
Here the gauge superfield $V^{++}$ and the hypermultiplet $q^+$ in the adjoint representation of the gauge group satisfy the analyticity conditions, and $F^{++}$ is the harmonic superspace analog of the gauge field strength.

In components the action (\ref{6D_Action}) (among others) contains the term with higher derivatives of the gauge field
\begin{equation}
S = \mbox{tr}\int d^6x\,\bigg\{\pm \frac{1}{e_0^2}({\cal D}_\mu F^{\mu\nu})^2 + \ldots\bigg\}.
\end{equation}

Due to the presence of higher derivatives, the degree of divergence for the theory (\ref{6D_Action}) does not increase with a number of loops. The possible divergences are either quadratic or logarithmical, but the quadratic divergences cancel each other in the one-loop approximation (and presumably in all loops). The theory (\ref{6D_Action}) is not anomalous \cite{Smilga:2006ax} and seems to be renormalizable. Moreover, the hypermultiplet and ghosts do not receive divergent quantum corrections \cite{Buchbinder:2025aac}.

To reveal possible features of the quantum correction structure, we regularize the theory (\ref{6D_Action}) by higher covariant derivatives. The higher derivative term is constructed with the help of the operator
\begin{equation}
\Box \equiv \frac{1}{2} (D^+)^4 (\nabla^{--})^2,
\end{equation}

\noindent
which is analogous to the Laplace operator when acting on analytic superfields. Then the regularized action can be written in the form
\begin{equation}
S_{\mbox{\tiny reg}} = \pm \frac{1}{2e_0^2} \mbox{tr} \int d\zeta^{(-4)} F^{++} R\Big(\frac{\Box}{\Lambda^2}\Big) F^{++} - \frac{2}{e_0^2}\,\mbox{tr} \int d\zeta^{(-4)} \widetilde{q^+} \nabla^{++} q^+,
\end{equation}

\noindent
where $R(0)=1$ and $R(x)\to \infty$ at $x\to \infty$. For regularizing the residual one-loop divergences it is also necessary to add the Pauli--Villars superfields with the mass $M=a \Lambda$ as discussed in detail in \cite{Buchbinder:2025aac}.

After calculating one-loop divergent superdiagrams, it was obtained that the quadratic divergences cancel each other, while the sum of the logarithmical ones gives the $\beta$-function $\beta(\alpha_0)$ of the form
\begin{equation}
\frac{\beta(\alpha_0)}{\alpha_0^2} = \mp 2\pi C_2\, \int \frac{d^6q}{(2\pi)^6} \frac{d}{d\ln\Lambda} \frac{\partial}{\partial q_\mu}\frac{\partial}{\partial q^\mu} \bigg[\frac{1}{q^4} \ln\Big(1+ \frac{M^4}{q^4 R(q^2/\Lambda^2)}\Big)\bigg] + O(\alpha_0).
\end{equation}

\noindent
Therefore, exactly as in the $4D$ case, the $\beta$-function is given by integrals of double total derivatives with respect to the loop momentum. Note that, due to the presence of an arbitrary regulator function $R(x)$, this fact is highly nontrivial. After calculating the loop integral we obtain the one-loop result
\begin{equation}
\beta(\alpha_0) = \mp \frac{\alpha_0^2 C_2}{2\pi^2} + O(\alpha_0^3).
\end{equation}

\noindent
This expression agrees with the results of the calculations made in \cite{Ivanov:2005qf,Casarin:2019aqw,Buchbinder:2020tnc,Buchbinder:2020ovf} by various methods after taking into account the contribution of the hypermultiplet $q^+$ in the adjoint representation.

The resemblance in the structure of the one-loop results for $4D$, ${\cal N}=1$ supersymmetric Yang--Mills theory and for the $6D$, ${\cal N}=(1,0)$ higher derivative theory under consideration allows suggesting that it may be possible to construct an all-loop exact expression for the $\beta$-function. In \cite{Buchbinder:2025aac} it was suggested that in the $6D$ case the result has the form
\begin{equation}\label{6D_Exact_Beta}
\beta(\alpha_0) = \mp \frac{\alpha_0^2 C_2}{2\pi^2\Big(1\mp \alpha_0 C_2/8\pi^2\Big)}.
\end{equation}

\noindent
Certainly, this guess should be verified by explicit multiloop calculations and (if possible) rigorously proved in all orders. If the expression (\ref{6D_Exact_Beta}) is really true, then after integrating the renormalization group equation we obtain that the expression
\begin{equation}\label{6D_RGI}
\Big(\frac{\alpha}{\mu^4}\Big)^{C_2}\exp\Big(\pm\frac{8\pi^2}{\alpha}\Big)=\mbox{RGI}
\end{equation}

\noindent
does not receive quantum corrections in any order of the perturbation theory and is a $6D$ analog of Eq. (\ref{N=1_SYM_RGI}).

\section*{Conclusion}
\hspace*{\parindent}

For certain ${\cal N}=1$ supersymmetric theories with multiple gauge couplings it is possible to construct such combinations of various couplings that do not depend on scale in all orders or, in other words, RGIs. In particular, in ${\cal N}=1$ SQCD interacting with ${\cal N}=1$ SQED such an RGI can be constructed from the strong and electromagnetic coupling constants (if the matter superfields have the same absolute values of the electromagnetic charges). This in particular implies that in this theory two gauge couplings do not run independently. For the MSSM (and also for NMMSM) one can construct two independent RGIs from the gauge couplings, Yukawa couplings and the $\mbox{\boldmath$\mu$}$ parameter. They are scale independent in all orders in the HD+MSL scheme, when a theory is regularized by Slavnov's higher covariant derivative method, and divergences are removed by minimal subtractions of logarithms. This fact has been confirmed by explicit calculations in the three-loop approximation, where the dependence on the renormalization prescription is already essential. However, in the $\overline{\mbox{DR}}$ scheme the above mentioned RGIs start to depend on scale from the three-loop order due to the scheme dependence.

We also argued that the behaviour of quantum corrections in a certain $6D$, ${\cal N}=(1,0)$ supersymmetric theory with higher derivatives is very similar to the one in the pure $4D$, ${\cal N}=1$ SYM theory. In particular, this theory presumably possesses an exact NSVZ-like $\beta$-function, which leads to the all-loop renormalization group invariance of the expression (\ref{6D_RGI}).


\end{document}